\documentstyle[ltwol]{article}

\def\PLB{{\em Phys. Lett.}  B}

\def\PRD{{\em Phys. Rev.} D}

\def\ra{\rightarrow}

\def\al{\alpha}

\def\be{\begin{equation}}
\def\ee{\end{equation}}
\def\bea{\begin{eqnarray}}
\def\eea{\end{eqnarray}}
\begin{document}

\title{NONLOCAL COLOR INTERACTIONS IN A GAUGE-INVARIANT FORMULATION OF QCD}

\author{Kurt Haller and Lusheng Chen}

\address{Department of Physics, University of Connecticut, 
Storrs, CT 06269\\E-mail: khaller@uconnvm.uconn.edu}     

%%%%%%%%%%%%%%%%% abstract starts here%%%%%%%%%%%%%%%%%%%%%%%%%%
\twocolumn[\maketitle\abstracts
{ We construct a set of states that implement the non-Abelian 
Gauss's law for QCD. We also construct a set of 
gauge-invariant operator-valued quark and gluon fields by establishing an explicit unitary 
equivalence between the Gauss's law operator and the
`pure glue' part of the Gauss's law operator. This unitary equivalence enables 
us to use the `pure glue' Gauss's law operator to represent the entire Gauss's law operator
in a new representation. Since the quark field commutes with the `pure glue' Gauss's law operator, 
it is a gauge-invariant 
field in this new representation. We use the unitary equivalence of the 
new and the conventional representations to construct gauge-invariant quark and gluon fields
in both representations,  
and to transform the QCD Hamiltonian in the temporal 
gauge so that it is expressed entirely in terms of gauge-invariant quantities.
In that form, 
all interactions between quark fields mediated by `pure gauge' components of the 
gluon field have been transformed
away, and replaced by a nonlocal interaction between gauge-invariant color-charge densities. This feature ---
that, in gauge-invariant formulations, interactions mediated by pure gauge components of gauge 
fields have been replaced by nonlocal interactions --- is shared by many gauge theories. In 
QED, the resulting nonlocal interaction is the Coulomb interaction, which is the 
Abelian analog of the QCD interaction we have identified and are describing in this work. 
The leading term, in a multipole expansion, 
of this nonlocal QCD interaction vanishes for quarks in color-singlet 
configurations, suggesting a dynamical origin for color confinement; 
higher order terms of this multipole expansion suggest a QCD mechanism for color 
transparency. We also 
show how, in an SU(2) model, this nonlocal interaction can be evaluated nonperturbatively.}]
%%%%%%%%%%%%%%%%% newsection - INTRO %%%%%%%%%%%%%%%%%%%%%%%%%%
\section{Introduction}\label{sec:int}
The program of the work we are discussing is designed to take advantage of a common thread that 
appears to underlie all gauge theories --- non-Abelian as well as Abelian: 
When only gauge-invariant fields are used in constructing the Hamiltonian for a gauge theory, 
interactions between charged fields  (electrical, color, etc.) and pure-gauge components of gauge fields 
cannot arise, and a nonlocal interaction between charge densities --- 
the Coulomb interaction in QED ---  appears in their stead.
Moreover, in gauge-invariant QED, 
the only interaction besides the Coulomb interaction is between the transverse (gauge-invariant) 
part of the gauge field and the current density; this interaction takes the form 
$-\,{\int}j_i({\bf r})A_{T\,i}\,({\bf r})d{\bf r}$, where $A_{T\,i}$ is the transverse (gauge-invariant) 
part of the gauge field, and 
$\vec{j}({\bf r})=e{\psi^\dagger}({\bf r})\vec{\alpha}\psi({\bf r})$. Since the current density has a $v/c$ 
dependence in the nonrelativistic limit, the Coulomb interaction is,
by far, the most important electrodynamic force in the low-energy regime. \smallskip

We will show that a very similar 
effect occurs in QCD. The main difference between the two cases is the fact that constructing gauge-invariant 
gauge fields is much more difficult in non-Abelian theories than in Abelian ones. Also, the nonlocal 
interaction in QCD is more complicated, and involves not only  quark color-charge densities, but also 
gauge-invariant gluon fields. 
Significantly, in QCD, just as in QED, the only interaction other than the nonlocal one 
involves the gauge-invariant gauge (gluon) field and a quark color-current density that is the non-Abelian 
analog of the current density in QED, and that can be expected to also have a $v/c$ 
dependence in the nonrelativistic limit.\smallskip

In this paper we will show how to implement the non-Abelian Gauss's law, how to construct 
gauge-invariant quark and gluon states, and how to express the QCD Hamiltonian in terms of these gauge-invariant 
fields. We will also discuss the physical implications of the formal results we obtain.  
%%%%%%%%%%%%%%%%% newsection IMPLEMENTING GAUSS %%%%%%%%%%%%%%%%%%%%%%%%%%
\section{Implementing Gauss's law for QCD}\label{sec:Gauss}
The Gauss's law operator in QCD is given by 
\be
{\hat {\cal G}}^a({\bf{r}})=\overbrace{\partial_i\Pi^a_i({\bf{r}})+
\underbrace{gf^{abc}A^b_i({\bf{r}}){\Pi}^c_i({\bf{r}})}_{J^a_0({\bf{r}})}}^
{D_i\Pi^a_i({\bf{r}})=\mbox{`pure-glue' part}}+j^a_0({\bf{r}}),  
\label{eq:allgauss}
\ee
where $j^a_0({\bf{r}})=g\psi^{\dagger}({\bf{r}})\frac{{\lambda}^a}{2} 
\psi({\bf{r}})$ is the quark color-charge density, and $J^a_0({\bf{r}})$ is the gluon
color-charge density. To implement the non-Abelian Gauss's law, we must construct states that 
are annihilated by the Gauss's law operator, $i.\,e.$ we must solve 
\begin{equation}
{\hat {\cal G}}^a({\bf{r}})\,|{\hat {\Psi}}{\rangle}=0\,.
\label{eq:impgauss}
\end{equation}
In QED, it is easy to implement Gauss's law, $\{\partial_{i}\Pi_{i}({\bf r})+j_{0}({\bf r})\}
\,|{\hat {\xi}}{\rangle}=0\,,$ because $\partial_{i}\Pi_{i}({\bf r})+j_{0}({\bf r})$ and 
$\partial_{i}\Pi_{i}({\bf r})$ are unitarily equivalent, and $\partial_{i}\Pi_{i}({\bf r})
\,|{\xi}{\rangle}=0$ is simple to solve.\cite{abelqed} In QCD, this method for implementing 
Gauss's law is not available. ${\hat {\cal G}}^a({\bf{r}})$ and $\partial_i\Pi^a_i({\bf{r}})$
cannot be unitarily equivalent, because these two operator-valued quantities obey 
inequivalent commutator algebras; whereas 
$\left[\partial_{i}\Pi_{i}^{a}({\bf r}), \partial_{i}\Pi_{i}^{b}({\bf r}^{\prime})\right]=0$, 
$\left[{\hat {\cal G}}^a({\bf{r}}),
{\hat {\cal G}}^b({\bf{r^\prime}})\right]=
igf^{abc}{\hat {\cal G}}^c({\bf{r}){\delta}({\bf r-r^\prime}})\,.$~\cite{khtemp} \smallskip

We will initially implement the `pure glue' form of Gauss's law for QCD, 
\be
D_i\Pi^a_i({\bf{r}})|{\Psi}\rangle=0\,,
\label{eq:pureQCD}
\ee
by constructing a state $|{\Psi}\rangle={\Psi}\,|{\phi}\rangle$, for which 
$D_i\Pi^a_i({\bf{r}})|{\Psi}\rangle=\{\,\partial_{i}\Pi^a_{i}({\bf{r}}) + J_{0}^{a}({\bf{r}})\,\}
{\Psi}\,|{\phi}\rangle =0$; $|{\phi}\rangle$ represents a state that is annihilated by 
$\partial_{i}\Pi^a_{i}({\bf{r}})$ --- the so-called `Fermi' state.\cite{fermi} We then seek 
to construct an operator ${\Psi}$, for which 
\be
[\,\partial_{i}\Pi^a_{i}({\bf{r}}),\,{\Psi}\,]=-J_{0}^{a}({\bf{r}})\,
{\Psi}\,+\,B_Q^{a}({\bf{r}})\,.
\label{eq:commpsi}
\ee 
where $B_{Q}^{a}({\bf{r}})$ is an 
operator that has
$\partial_{i}\Pi^a_{i}({\bf{r}})$ on its extreme right.
Eq.(\ref{eq:commpsi}) is essentially an operator differential equation, in which the 
commutator, $[\,\partial_{i}\Pi^a_{i}({\bf{r}}),\,{\Psi}\,]$, is a derivative. \smallskip

We have found a solution of this equation,\cite{BCH1} in which $\Psi$ is expressed as
\be
{\Psi} =||\exp({\cal{A}})\,||
\label{eq:defpsi}
\ee
with
\be
{\cal{A}}=i{\int}d{\bf{r}}\;\overline{{\cal{A}}_{i}^{\gamma}}({\bf{r}})\;
\Pi_i^{\gamma}({\bf{r}})\,,
\label{eq:scriptA}
\ee
 and with $\overline{{\cal{A}}_{i}^{\gamma}}({\bf{r}})$ represented as the 
series 
\be
\overline{{\cal{A}}_{i}^{\gamma}}({\bf{r}})=\sum_{n=1}^\infty g^n
{\cal{A}}_{(n)i}^{\gamma}({\bf{r}})\;. 
\label{eq:expandA}
\ee
The ${\cal{A}}_{(n)i}^{\gamma}({\bf{r}})$ are nonlinear functionals of transverse and 
longitudinal parts of gauge fields, but are independent of the canonical momentum $\Pi_i^{\gamma}({\bf{r}})$.
The ordered product ${\|}\,\exp({\cal{A}})\,{\|}$  is defined so that, in the $n^{th}$ order 
term, ${\|}({\cal{A}})^n\,{\|}\;,$ all 
functionals of the gauge field $A^a_i$ are {\em to the left of} all functionals of the 
canonical momenta $\Pi^b_j\,.$ 
We refer to $\overline{{\cal{A}}_{i}^{\gamma}}({\bf{r}})$ as the {\em resolvent gauge field.}\smallskip

The requirement that $|\Psi\rangle$ implement Gauss's law can be translated into a condition on the 
resolvent gauge field. Before we formulate this condition, we first define the following quantities:\smallskip

${\cal{X}}^\alpha({\bf{r}})=\,{\textstyle\frac{\partial_j}{\partial^2}}A_j^\alpha({\bf{r}})$ and
 ${\cal{R}}^{\vec{\alpha}}_{(\eta)}({\bf{r}})=\prod_{m=1}^\eta{\cal{X}}^{\alpha[m]}({\bf{r}})$, 
which are functionals of gauge fields; 
$\overline{{\cal Y}^{\alpha}}({\bf r})=
{\textstyle \frac{\partial_{j}}{\partial^{2}}\overline{{\cal A}_{j}^{\alpha}}({\bf r})}$ and
${\cal{M}}_{(\eta)}^{\vec{\alpha}}({\bf{r}})=\!\!\prod_{m=1}^\eta
\overline{{\cal Y}^{\alpha[m]}}({\bf{r}})$,
which have structures similar to ${\cal{X}}^\alpha({\bf{r}})$ 
and ${\cal{R}}^{\vec{\alpha}}_{(\eta)}({\bf{r}})$ respectively, 
 but which are functionals of the resolvent gauge fields. 
We also need to define:
\begin{eqnarray}
f^{\vec{\alpha}\beta\gamma}_{(\eta)}=&&f^{\alpha[1]\beta b[1]}\,
\,f^{b[1]\alpha[2]b[2]}\,f^{b[2]\alpha[3]b[3]}\,\times\cdots \nonumber \\
&&{\times}f^{b[\eta-2]\alpha[\eta-1]b[\eta-1]}f^{b[\eta-
1]\alpha[\eta]\gamma}\;;
\label{eq:fchain}
\end{eqnarray} 
these are chains of structure constants whose `links' are summed over repeated indices.

The condition on the resolvent gauge field that is equivalent to implementing the `pure glue' Gauss's law, is
\begin{eqnarray}
&&igf^{a\beta d}A_{i}^{\beta}({\bf r})\int d{\bf r}^\prime
[\,\Pi_{i}^{d}({\bf r}),\,
\overline{{\cal A}_{j}^{\gamma}}({\bf r}^\prime)\,]\,
V_{j}^{\gamma}({\bf r}^\prime)\,+\nonumber\\ 
&&i\int d{\bf r}^\prime[\,\partial_{i}\Pi_{i}^{a}({\bf r}),\,
\overline{{\cal A}_{j}^{\gamma}}({\bf r}^\prime)\,]\,
V_{j}^{\gamma}({\bf r}^\prime)
+gf^{a\mu d}\,A_{i}^{\mu}({\bf r})\,V_{i}^{d}({\bf r})\,
\nonumber \\ 
&&=\sum_{\eta=1}{\textstyle\frac{g^{\eta +1}B(\eta)}{\eta!}}\,
f^{a\beta c}f^{\vec{\alpha}c\gamma}_{(\eta)}\,A_{i}^{\beta}({\bf r})\,
{\textstyle \frac{\partial_{i}}{\partial^{2}}}\left(\,
{\cal M}_{(\eta)}^{\vec{\alpha}}({\bf r})\,
\partial_{j}V_{j}^{\gamma}({\bf r})\,\right)\,-
\nonumber \\ 
&&\sum_{\eta=0}\sum_{t=1} (-1)^{t-1}g^{t+\eta}{\textstyle \frac{B(\eta)}{\eta!(t-1)!(t+1)}}\,\,\times\,\,\nonumber\\
&&\;\;\;\;\;\;\;\;\;\;\;\;\;\;\;f^{\vec{\mu}a\lambda}_{(t)}f^{\vec{\alpha}\lambda\gamma}_{(\eta)}\,
{\cal R}_{(t)}^{\vec{\mu}}({\bf r})\,
{\cal M}_{(\eta)}^{\vec{\alpha}}({\bf r})\,
\partial_{i}V_{i}^{\gamma}({\bf r})
 \nonumber\\ 
&&-gf^{a\beta d}A_{i}^{\beta}({\bf r})\,
\sum_{\eta=0}\sum_{t=1} (-1)^{t}g^{t+\eta}\,
{\textstyle\frac{B(\eta)}{\eta!(t+1)!}}\times \nonumber\\ 
&&\;\;\;\;\;\;\;\;\;\;\;\;\;f^{\vec{\mu}d\lambda}_{(t)}
f^{\vec{\alpha}\lambda\gamma}_{(\eta)}
{\textstyle\frac{\partial_{i}}{\partial^{2}}}\,
\left(\,{\cal R}_{(t)}^{\vec{\mu}}({\bf r})\,
{\cal M}_{(\eta)}^{\vec{\alpha}}({\bf r})\,
\partial_{j}V_{j}^{\gamma}({\bf r})\,\right),
\label{eq:condA}
\end{eqnarray}
where $B({\eta})$ is the ${\eta}^{th}$ Bernoulli number, and $V_{j}^{\gamma}({\bf r})$
represents any arbitrary vector field in the adjoint representation of SU(3). \smallskip

We have solved Eq.(\ref{eq:condA});~\cite{CBH2} The solution is given by  
\bea
&&{\int}d{\bf r}\overline{{\cal A}_{j}^{\gamma}}({\bf r})V_{j}^{\gamma}({\bf r})=
\sum_{\eta=1}^\infty
{\textstyle\frac{ig^\eta}{\eta!}}{\int}d{\bf r}\;
\!\left\{\,\psi^{\gamma}_{(\eta)j}({\bf{r}})\,+\right.\nonumber\\
&&\left.\;\;\;\;\;\;\;\;\;\;\;\;\;\;\;f^{\vec{\alpha}\beta\gamma}_{(\eta)}\,
{\cal{M}}_{(\eta)}^{\vec{\alpha}}({\bf{r}})\,
\overline{{\cal{B}}_{(\eta) j}^{\beta}}({\bf{r}})\,\right\}\;
\!\!V_{j}^{\gamma}({\bf r})\;,
\label{eq:inteq2}
\eea 
where 
\be
\psi^{\gamma}_{(\eta)i}({\bf{r}})= \,(-1)^{\eta-1}\,
f^{\vec{\alpha}\beta\gamma}_{(\eta)}\,
{\cal{R}}^{\vec{\alpha}}_{(\eta)}({\bf{r}})\;
{\cal{Q}}_{(\eta)i}^{\beta}({\bf{r}})
\label{eq:psi}
\ee
\be
\mbox{with}\;\;{\cal{Q}}_{(\eta)i}^{\beta}({\bf{r}}) =
[\,a_i^\beta ({\bf{r}})+
{\textstyle\frac{\eta}{(\eta+1)}}\,x_i^\beta({\bf{r}})\,]\;,
\label{eq:calQ}
\ee and
\be
\overline{{\cal B}_{(\eta) i}^{\beta}}({\bf r})=
a_i^{\beta}({\bf r})+\,
\left(\,\delta_{ij}-{\textstyle\frac{\eta}{(\eta+1)}}
{\textstyle\frac{\partial_{i}\partial_{j}}{\partial^{2}}}\,\right)
\overline{{\cal A}_{j}^{\beta}}({\bf r})\,;
\label{eq:calB}
\ee
$a_i^\al$ and $x_i^\al$ designate the transverse and longitudinal parts of 
the gauge field $A_i^\al ({\bf r})$ respectively. Eq.(\ref{eq:inteq2}) calls
for summations over the multiplicity index ${\eta}$, which labels the multiplicity of 
$\overline{{\cal Y}^{\alpha}}({\bf r})$ factors in ${\cal{M}}_{(\eta)}^{\vec{\alpha}}({\bf{r}})$, 
and the multiplicity of ${\cal{X}}^\alpha({\bf{r}})$ factors in ${\cal{R}}^{\vec{\alpha}}_{(\eta)}({\bf{r}})$.
This multiplicity of factors makes Eq.(\ref{eq:inteq2}) a 
nonlinear integral equation that specifies the resolvent gauge field
$\overline{{\cal A}_{j}^{\gamma}}({\bf r})$ recursively. The possibility that 
this nonlinear equation has multiple solutions, and that these multiple solutions correspond to 
different physical regimes, is interesting, but as yet unexplored. \smallskip

In our earlier work,~\cite{CBH2} we proved that Eq.(\ref{eq:inteq2}) is a solution of 
Eq.(\ref{eq:condA}) --- a proof that we have referred to as the `fundamental theorem' --- 
 and that the resolvent gauge field, therefore, is the operator-valued
field required for the implementation of the non-Abelian Gauss's law. We will see, 
in the remainder of this paper, that the resolvent gauge field plays a pivotal role in 
the construction of gauge-invariant fields and in the transformation of the QCD Hamiltonian 
to a functional of gauge-invariant fields.  
%%%%%%%%%%%%%%%%% newsection GAUGE_INV_FIELDS %%%%%%%%%%%%%%%%%%%%%%%%%%
\section{Gauge-invariant quark and gluon fields}\label{sec:GI}
The observation that underlies the construction of gauge-invariant quark and gluon fields is that 
the Gauss's law operator, ${\hat {\cal G}}^a({\bf{r}})$, and the `pure glue' Gauss' law operator 
${\cal G}^a({\bf{r}})=D_i\Pi^a_i({\bf{r}})$ are unitarily equivalent. In earlier work,~\cite{CBH2} 
we have shown that 
\begin{equation}
{\hat {\cal G}}^{a}({\bf r})
={\cal{U}}_{\cal{C}}\,
{\cal G}^{a}({\bf r})\,{\cal{U}}^{-1}_{\cal{C}}\,,
\label{eq:Gtrans}
\end{equation} 
where ${\cal{U}}_{\cal{C}}=\exp({\cal C}_{0})
\exp({\bar {\cal C}})$ with
${{\cal C}_{0}}$ and ${\bar {\cal C}}$ given by
\be
{\cal C}_{0}=i\,\int d{\bf{r}}\,
{\textstyle {\cal X}^{\alpha}}({\bf r})\,j_{0}^{\alpha}({\bf r})\;,
\;\;\;\;\;\mbox{and}
\ee
\be
{\bar {\cal C}}=i\,\int d{\bf{r}}\,
\overline{{\cal Y}^{\alpha}}({\bf r})\,j_{0}^{\alpha}({\bf r})\;.
\label{eq:CCbar}
\ee
We are therefore free to interpret the `pure glue' Gauss's law operator ${\cal G}^{a}({\bf r})$
as the complete Gauss's law operator, ${\hat {\cal G}}^a({\bf{r}})$, in a different, unitarily 
equivalent representation. We will refer to the conventional representation, in which 
${\hat {\cal G}}^a({\bf{r}})$ is the complete Gauss's law operator and ${\cal G}^{a}({\bf r})$
the `pure glue' Gauss's law operator, as the ${\cal C}$ representation; and the new 
 representation, in which ${\cal G}^{a}({\bf r})$ is the unitarily transformed {\em complete} 
Gauss's law operator, as the ${\cal N}$ representation. It is manifest that the spinor (quark) field 
$\psi({\bf r})$ commutes with ${\cal G}^{a}({\bf r})$. Since the Gauss's law operator is the 
generator of gauge transformations, $\psi({\bf r})$ is a gauge-invariant spinor (quark) field in the ${\cal N}$ 
representation. To find the corresponding gauge-invariant quark field in the ${\cal C}$ representation, we 
apply the transformation that appears in Eq.(\ref{eq:Gtrans}), and obtain 
\be
{\psi}_{\sf GI}({\bf{r}})={\cal{U}}_{\cal C}\,\psi({\bf{r}})\,{\cal{U}}^{-1}_{\cal C}=
V_{\cal{C}}({\bf{r}}){\psi}({\bf{r}})
\label{eq:psiGI}
\end{equation}
where
\begin{equation}
V_{\cal{C}}({\bf{r}})=
\exp\left(\,-ig{\overline{{\cal{Y}}^\alpha}}({\bf{r}})
{\textstyle\frac{\lambda^\alpha}{2}}\,\right)\,
\exp\left(-ig{\cal X}^\alpha({\bf{r}})
{\textstyle\frac{\lambda^\alpha}{2}}\right)\;,
\label{eq:el1}
\end{equation}
and the ${\lambda}^h$ represent the Gell-Mann SU(3) matrices.
$V_{\cal{C}}({\bf{r}})$ could be written as $\exp(-ig{\cal Z}^{\al}({\bf{r}})
{\textstyle\frac{\lambda^\alpha}{2}})$,
where the Baker-Hausdorff-Campbell theorem can be used to express ${\cal Z}^{\al}$ as a functional 
of ${\cal X}^\alpha$ and ${\overline{{\cal{Y}}^\alpha}}$. 
$V_{\cal{C}}({\bf{r}})$ takes the form of a unitary operator that carries out a 
gauge transformation on the spinor field $\psi({\bf r})$; but, in this case, ${\cal Z}^{\al}$ is not a c-number
function in the adjoint representation of SU(3), but is a complicated functional of the gauge field. 
Under a gauge transformation, precisely compensating transformations are made on $\psi({\bf r})$ and
$V_{\cal{C}}({\bf{r}})$, so that ${\psi}_{\sf GI}({\bf{r}})$ is strictly gauge-invariant. When they appear 
In the ${\cal N}$ representation,  the color-charge and color-current densities, 
$j^a_0({\bf{r}})=g\psi^{\dagger}({\bf{r}})\frac{{\lambda}^a}{2} \psi({\bf{r}})$ and 
$j^a_i({\bf{r}})=g\psi^{\dagger}({\bf{r}}){\al}_i\frac{{\lambda}^a}{2} \psi({\bf{r}})$ 
respectively, therefore are gauge-invariant, although, in the ${\cal C}$ representation, both
of these quantities transform gauge-{\em covariantly}, as vectors in the adjoint representation 
of SU(3). In the ${\cal N}$ representation, 
the quark field $\psi({\bf r})$
implicitly includes enough of the gluon field to achieve this gauge invariance. \smallskip

We have used Eq.(\ref{eq:psiGI}) to find the gauge-invariant gluon field 
\begin{equation}
[\,A_{{\sf GI}\,i}^{b}({\bf{r}})\,{\textstyle\frac{\lambda^b}{2}}\,]
=V_{\cal{C}}({\bf{r}})\,[\,A_{i}^b({\bf{r}})\,
{\textstyle\frac{\lambda^b}{2}}\,]\,
V_{\cal{C}}^{-1}({\bf{r}})
+{\textstyle\frac{i}{g}}\,V_{\cal{C}}({\bf{r}})\,
\partial_{i}V_{\cal{C}}^{-1}({\bf{r}})\;,
\label{eq:gigf}
\end{equation}
or, equivalently,~\cite{CBH2}
\begin{equation}
A_{{\sf GI}\,i}^b({\bf{r}})=
A_{T\,i}^b({\bf{r}}) +
[\delta_{ij}-{\textstyle\frac{\partial_{i}\partial_j}
{\partial^2}}]\overline{{\cal{A}}^b_{j}}({\bf{r}})\,.
\label{eq:Adressedthree1b}
\end{equation}
We observe that, as in QED, the gauge-invariant gauge field $A_{{\sf GI}\,i}^b({\bf{r}})$ 
is transverse. But it is not merely the 
transverse part of the gauge field. In contrast to the gauge-invariant gauge field in QED, 
$A_{{\sf GI}\,i}^b({\bf{r}})$ also involves the transverse part of the resolvent gauge field. \smallskip

We have expanded ${\psi}_{\sf GI}({\bf{r}})$ and $A_{{\sf GI}\,i}^b({\bf{r}})$, 
and have verified that our gauge-invariant 
quark and gluon fields agree with the perturbative calculations of Lavelle, McMullan, $et.\,al.$ to 
the highest order to which their perturbative calculations were available.~\cite{lavelle2,lavelle3}
%%%%%%%%%%%%%%%%% newsection - GIHAM %%%%%%%%%%%%%%%%%%%%%%%%%%
\section{A gauge-invariant QCD Hamiltonian}%\label{sec:int} 
We have been able to express the QCD Hamiltonian 
\begin{eqnarray}
H_{\cal QCD} = \int d{\bf r} \left\{\ {\textstyle \frac{1}{2}
\Pi^{a}_{i}({\bf r})\Pi^{a}_{i}({\bf r})
+ \frac{1}{4}} F_{ij}^{a}({\bf r}) F_{ij}^{a}({\bf r})\,+\right.
\nonumber\\
 \left.{\psi^\dagger}({\bf r})
\left[\,{\beta} m-i\alpha_{i}
\left(\,\partial_{i}-igA_{i}^{a}({\bf r})
{\textstyle\frac{\lambda^\alpha}{2}}\,\right)\,\right]
\psi({\bf r})\right\}\,,
\label{eq:QCDHAM}
\end{eqnarray}
entirely in terms of gauge-invariant quark and gluon fields,\cite{BCH3} by systematically
transforming it, term by term, from the ${\cal C}$ to the ${\cal N}$ representation. Under this 
transformation, $\psi({\bf r}){\ra}V^{-1}_{\cal{C}}({\bf{r}})\psi({\bf r})$,
$\psi^{\dagger}({\bf r}){\ra}\psi^{\dagger}({\bf r})V_{\cal{C}}({\bf{r}})$, but the gauge field 
remains untransformed. Shifting $V_{\cal{C}}({\bf{r}})$ to the right and 
$V^{-1}_{\cal{C}}({\bf{r}})$ to the left until they encounter the $A_{i}^{a}({\bf r})
{\textstyle\frac{\lambda^\alpha}{2}}$, turns the second line of Eq.(\ref{eq:QCDHAM}) into 
${\int}d{\bf r}\,{\psi^\dagger}({\bf r})\left(\beta m-i\alpha_{i}
\partial_{i}\right)\psi({\bf r}) + \tilde{H}_{j-A}$, with 
\begin{equation}
\tilde{H}_{j-A}=-\,\int d{\bf r}\,\overbrace{g\,{\psi^\dagger}({\bf r})\alpha_{i}
{\textstyle\frac{\lambda^h}{2}}\psi({\bf r})}^{j^h_i({\bf r})}\,
A_{{\sf GI}\,i}^{h}({\bf r})\,.
 \label{eq:HJA}
\end{equation}
$ F_{ij}^{a}({\bf r}) F_{ij}^{a}({\bf r})$ remains untransformed in this process, but we found that 
this similarity transformation, when applied to $\Pi^{a}_{i}({\bf r})\Pi^{a}_{i}({\bf r})$, 
generates nonlocal interactions between color-charge densities
which we will discuss below.\cite{BCH3}\smallskip

The QCD Hamiltonian that results from the transformation to the ${\cal N}$ representation is 
\begin{eqnarray}
\tilde{H}_{\cal QCD} = &&\!\!\!\int d{\bf r} \left\{\ {\textstyle \frac{1}{2}
\Pi^{a}_{i}({\bf r})\Pi^{a}_{i}({\bf r})
+ \frac{1}{4}} F_{ij}^{a}({\bf r}) F_{ij}^{a}({\bf r})\,+\right.
\nonumber\\
 &&\left.{\psi^\dagger}({\bf r})
\left[\,{\beta} m-i\alpha_{i}
\,\partial_{i}\,\right]
\psi({\bf r})\right\}+ {\tilde {H}}^{\prime}\,,
\label{eq:NHAM}
\end{eqnarray}
where $\tilde{H}^{\prime}$ describes interactions involving the gauge-invariant quark field. The 
parts of $\tilde{H}^{\prime}$ relevant to the dynamics of quarks and gluons can be expressed as 
\begin{equation}
\tilde{H}^{\prime}=\tilde{H}_{j-A}+\tilde{H}_{LR}\,,
\label{eq:Hprime}
\end{equation} 
where ${\tilde{H}}_{LR}$ is the nonlocal interaction 
\begin{equation}
{\tilde{H}}_{LR}=H_{g-Q}+H_{Q-Q}\,.
\label{eq:HLR}
\end{equation}
We will initially focus our attention on  
$H_{Q-Q}$ in this report, since it illustrates the most important features of our work. A useful 
formulation of $H_{Q-Q}$ can be given as~\cite{CH1}
\begin{equation}
H_{Q-Q}=\frac{1}{2}{\int}d{\bf r}d{\bf x}\,{j}_0^b({\bf r}){\cal F}^{ba}({\bf r},{\bf x}){j}_0^a({\bf x})
\label{eq:HQQalt}.
\end{equation}
where the Green's function ${\cal F}^{ba}({\bf r},{\bf x})$ is represented as
$$\!\!\!\!\!\!\!\!\!\!\!\!\!\!\!\!\!\!\!\!\!\!\!\!\!\!\!\!\!\!\!\!\!\!\!\!\!\!\!\!\!\!\!\!\!
{\cal F}^{ba}({\bf r},{\bf x})=\frac{{\delta}_{ab}}{4{\pi}|{\bf r}-{\bf x}|}\;+\nonumber$$
$$2g\,f^{{\delta}_{(1)}ba}\int\frac{d{\bf y}}{4{\pi}|{\bf r}-{\bf y}|}{A_{{\sf GI}\,i}^
{{\delta}_{(1)}}({\bf{y}})\,\partial_{i}\frac{1}{4{\pi}|{\bf y}-{\bf x}|}}\,+\nonumber $$
$$\cdots \;\;\;\;\; \;\;\;\;\; \;\;\;\;\; + \;\;\;\;\; \;\;\;\;\; \;\;\;\;\;\cdots 
\;\;\;\;\; \;\;\;\;\; \;\;\;\;\;+ \nonumber\\$$
$$(-1)^{(n-1)}(n+1)g^nf^{{\delta}_{(1)}bs_{(1)}}f^{s_{(1)}{\delta}_{(2)}s_{(2)}}
{\cdots}f^{s_{(n-1)}{\delta}_{(n)}a}{\times}$$
$$\int\frac{d{\bf y}_1}{4{\pi}|{\bf r}-{\bf y}_1|}
{A}_{{\sf GI}\,i}^{{\delta}_{(1)}}({\bf{y}_1})\,\partial_{i}
\int\frac{d{\bf y}_2}{4{\pi}|{\bf y}_1-{\bf y}_2|}\,\times\nonumber $$
$${A}_{{\sf GI}\,j}^{{\delta}_{(2)}}
({\bf y}_{2})\,\partial_{j}\int\frac{d{\bf y}_3}{4{\pi}|{\bf y}_2-{\bf y}_3|}\,
\cdots\int\frac{d{\bf y}_n}{4{\pi}|{\bf y}_{(n-1)}-{\bf y}_n|}{\times}\nonumber$$
\be
{A}_{{\sf GI}\,{\ell}}^{{\delta}_{(n)}}
({\bf y}_{n})\,\partial_{\ell}
\frac{1}{4{\pi}|{\bf y}_n-{\bf x}|}+\cdots
\label{eq:Fgreen}
\ee
We make the following observations about ${\cal F}^{ba}({\bf r},{\bf x})$: 
The initial term resembles the Coulomb interaction. The infinite series of further terms consists of chains 
through which the interaction is transmitted from one 
color-charge density to the other. Each
chain contains a succession of `links', which have the characteristic form
\be
\mbox{link}=gf^{s_{(1)}{\delta}s_{(2)}}
{A}_{{\sf GI}\,j}^{{\delta}}
({\bf x})\,\partial_{j}\frac{1}{4{\pi}|{\bf x}-{\bf y}|}\,.
\label{eq:link}
\ee 
The `links' are coupled through summations over the $s_{(n)}$ indices and 
integrations over the spatial variables. As noted before, 
all the quantities in $H_{Q-Q}$ are gauge-invariant --- the color-charge density 
${j}_0^b({\bf r})$ as well as the gauge field ${A}_{{\sf GI}\,j}^{{\delta}}({\bf x})$.
$H_{g-Q}$ --- the other nonlocal interaction in ${\tilde{H}}_{LR}$ ---
couples quark to gluon color charge density. 
In $H_{g-Q}$ the quark color-charge density is coupled, through the same Green's function 
${\cal F}^{ba}({\bf r},{\bf x})$, to a gauge-invariant expression describing `glue'-color, 
which we have found to be~\cite{BCH3,CH1}  $${\sf K}_g^d({\bf r})=
gf^{d\sigma e}\,{\sf Tr}\left[V_{\cal{C}}^{-1}({\bf{r}})
{\textstyle\frac{\lambda^e}{2}}V_{\cal{C}}({\bf{r}}){\textstyle\frac{\lambda^b}{2}}\right]
A_{{\sf GI}\,i}^{\sigma}({\bf r})
\Pi^{b}_{i}({\bf r})\,.$$

The chains that constitute  ${\cal F}^{ba}({\bf r},{\bf x})$ --- coupled links, in which the $n^{th}$ 
order chain is a product of $n$ 
links --- suggest features closely associated with QCD: flux tubes, `string'-like structures tying 
colored objects to each other, etc. But these rudimentary analogies only serve to direct our attention to 
potentially important features of this interaction --- they do not, by themselves, demonstrate a 
physical effect. \smallskip

With regard to $\tilde{H}_{j-A}$, we observe that the color-current density $j^h_i({\bf r})$ has a 
configuration-space structure that is similar to  that of the electric current density in QED. We can 
therefore reasonably expect that it, too, will manifest a $v/c$ dependence for nonrelativistic quarks interacting 
with the purely transverse ${A}_{{\sf GI}\,j}^{{\delta}}({\bf x})$, and 
that ${\tilde{H}}_{LR}$ will be of predominant importance in the low-energy QCD regime.
%%%%%%%%%%%%%%%%% newsection - COLOR %%%%%%%%%%%%%%%%%%%%%%%%%%
\section{Quark confinement and color transparency}\label{sec:color}
The use of Eqs.(\ref{eq:HQQalt}) and (\ref{eq:Fgreen}) to explicitly evaluate an effective `potential' between quark 
color-charge densities is precluded, at the present time,  by the fact that we 
have not yet found an expression for 
the gauge-invariant gauge field ${A}_{{\sf GI}\,j}^{{\delta}}({\bf x})$. However, 
we can draw some important conclusions from the general form of Eqs.(\ref{eq:HQQalt}) 
and (\ref{eq:Fgreen}). \smallskip

If we assume that the Green's function ${\cal F}^{ba}({\bf r},{\bf x})$ varies smoothly as a function of $\bf x$, 
when evaluated in an appropriately chosen state, and that quarks, or configurations of quarks, are 
localized in wave packets whose size is small enough so that ${\cal F}^{ba}({\bf r},{\bf x})$ varies only moderately
over the space occupied by these wave packets, then we are entitled to make a `color-multipole' 
expansion of the expression
${\int}{\cal F}^{ba}({\bf r},{\bf x}){j}_0^a({\bf x})d{\bf x}$ --- which appears 
in Eq.(\ref{eq:HQQalt}) ---about the point ${\bf x}={\bf x}_0$,
in the form
\bea 
{\int}d{\bf x}\,&&\!\!\left\{{\cal F}^{ba}({\bf r},{\bf x}_0)+X_i\partial_i
{\cal F}^{ba}({\bf r},{\bf x}_0)+\nonumber\right.\\
&&\left.\frac{1}{2}\,X_iX_j\,\partial_i\partial_j
{\cal F}^{ba}({\bf r},{\bf x}_0)\;+\;\;\cdots\right\}{j}_0^a({\bf x})
\label{eq:Lmultipole}
\eea
where $X_i=(x-x_0)_i$ and $\partial_i=\partial /\partial x_i\,.$ When we perform the integration in 
Eq.~(\ref{eq:Lmultipole}), the first term contributes ${\cal F}^{ba}({\bf r},{\bf x}_0)\,
{\cal Q}^a\,,$ where 
${\cal Q}^a=\int\,d{\bf x}\,{j}_0^a({\bf x})$ (the integrated ``color charge''). 
Since the color charge is the 
generator of rotations in SU(3) space, 
it will annihilate any multi-quark state vector in a singlet color configuration. 
Multi-quark packets in a singlet color configuration therefore are immune to 
the leading term of the nonlocal 
$H_{Q-Q}$. Color-singlet configurations of quarks are only subject to the color multipole terms, 
which act as color analogs to the Van der Waals interaction.
The scenario that this model suggests is that the leading term in $H_{Q-Q}$, namely 
${\cal Q}^b{\cal F}^{ba}({\bf r}_0,{\bf x}_0){\cal Q}^a$ for a quark color charge ${\cal Q}^a$ at ${\bf r}_0$ and 
another quark color charge ${\cal Q}^b$ at ${\bf x}_0$, may be responsible for the confinement of 
quarks and packets of quarks that are not in color-singlet configurations. 
Moreover, assuming that ${\cal F}^{ba}({\bf r},{\bf x}_0)$ 
varies only gradually within a volume occupied by quark packets, the effect of the higher order
color multipole forces on a packet of quarks in a color-singlet configuration becomes more significant as 
the packet increases in size. As small quark packets move through gluonic matter, they 
will experience only insignificant effects from the multipole contributions to $H_{Q-Q}$, 
since, as can be seen from 
Eq.~(\ref{eq:Lmultipole}), the factors $X_i\,$, $X_iX_j$, ${\cdots}$, $X_{i(1)}{\cdots}X_{i(n)}$,
keep the higher order multipole terms from making significant contributions to 
${\int}d{\bf x}\,{\cal F}^{ba}({\bf r},{\bf x}){j}_0^a({\bf x})$ when they are integrated 
over small packets of quarks. 
As the size of the quark packets increases, the regions
over which the multipoles are integrated also increases, and the effect of the multipole interactions
on the color-singlet packets can become larger. 
This dependence on packet size of the final-state interactions experienced by color-singlet states ---
$i.\,e.$ the increasing importance of final-state interactions as color-singlet packets grow in size --- 
is in qualitative agreement with the characterizations of color transparency 
given by Miller and by Jain, Pire and Ralston.~\cite{miller} \smallskip

In making this argument, the following features of this gauge-invariant formulation must be borne in 
mind: The state vectors that can appropriately be chosen for such a representation must implement the 
non-Abelian Gauss's law applicable to QCD --- $i.\,e.$ the states would have to be annihilated by the Gauss's 
law operator, which is ${\cal G}$ in the ${\cal N}$ representation. That still 
allows us to apply quark creation operators 
to such states to form multi-quark Fock states with orbitals whose configuration-space dependence is arbitrary 
as far as the gauge problem is concerned. These states are far from being Fock states in the gluon sector --- 
they have to be complex enough to implement Gauss's law. But that fact does not inhibit us in using Fock 
states for multi-quark configurations, because, in the ${\cal N}$ representation, the quark fields contain
enough gluon contributions to be gauge-invariant by themselves. The multi-quark states therefore have a 
status similar to multi-electron states in a gauge-invariant formulation of QED ($e.\,g.$ in the 
Coulomb gauge), which are 
surrounded by a Coulomb field and therefore are gauge-invariant.

In order to test these ideas quantitatively, we have made use  
of Yang-Mills theory --- the SU(2) version of this model --- for which the structure constants 
$f^{{\delta}ba}$ are ${\epsilon}^{{\delta}ba}$.~\cite{CH1} 
We model $A_{{\sf GI}\,i}^{\delta}({\bf{r}})$, which, in this case, are transverse fields in the adjoint 
representation of SU(2), as the manifestly transverse ``hedgehog'' configuration
\begin{equation}
{\langle}A_{{\sf GI}\,i}^{\delta}({\bf{r}})\rangle={\epsilon}^{ij{\delta}}r_j{\phi}(r).
\label{eq:GISU2}
\end{equation}
Although there is no reason to believe that the {\em ansatz} given in Eq.~(\ref{eq:GISU2}) follows 
from the dynamical equations that determine $A_{{\sf GI}\,i}^{\delta}({\bf{r}})$, it is a convenient choice 
for examining to what extent the structure of ${\cal F}^{ba}({\bf r},{\bf x})$ --- 
as distinct from the precise form of $A_{{\sf GI}\,i}^{\delta}({\bf{r}})$ ---                                  
enables us to nonperturbatively evaluate the infinite series given in Eq.~(\ref{eq:Fgreen}). 
As we were able to show,\cite{CH1} the perturbative evaluation of $H_{Q-Q}$ for this SU(2) 
model could be bypassed, and the entire series in ${\cal F}^{ba}({\bf r},{\bf x})$ obtained from the 
solution of a sixth-order differential equation.

%%%%%%%%%%%%%%%%% newsection ACK %%%%%%%%%%%%%%%%%%%%%%%%%%
\section*{Acknowledgments}This research was supported by the Department of Energy
under Grant No. DE-FG02-92ER40716.00.
%%%%%%%%%%%%%%%%% newsection REF %%%%%%%%%%%%%%%%%%%%%%%%%%


\section*{References}
\begin{thebibliography}{99}
\bibitem{abelqed}P. A. M. Dirac, {\em Can. J. Phys.} 33, 650 (1955); \\K. Haller and E. Lim-Lombridas,
{\em Found. of Phys.} 24, 217 (1994), and further references cited therein.
\bibitem{khtemp}for example, R. Jackiw, {\em Rev. Mod. Phys.} 52, 661 (1980); K. Haller, \PRD 36, 1839 (1987). 
\bibitem{fermi}E. Fermi, {\em Rev. Mod. Phys.} 4, 661 (1932).
\bibitem{BCH1} M. Belloni, L. Chen and K. Haller, \PLB 373, 185 (1996).
\bibitem{CBH2}L. Chen, M. Belloni and K. Haller, \PRD 55, 2347 (1997).
\bibitem{lavelle2}M. Lavelle and D. McMullan, \PLB 329 (1994) 68.
\bibitem{lavelle3}E. Bagan, M. Lavelle, B. Fiol, N. Roy and D. McMullan, 
{\em Constituent Quarks from QCD: Perturbation Theory and the Infrared.}
hep-ph/9609330. 
\bibitem{BCH3} M. Belloni, L. Chen and K. Haller, \PLB 403, 316 (1997).
\bibitem{CH1} L. Chen and K. Haller, {\em Quark confinement and color transparency 
in a gauge-invariant formulation of QCD,} hep-th/9803250.
\bibitem{miller}G. A. Miller, {\em Color Transparency --- Color Coherent Effects in Nuclear Physics.}
DOE/ER/41014-26-N97; nucl-th/9707040; P. Jain, B. Pire, and J. P. Ralston, {\em Physics Reports} 271, 67 (1996).
\end{thebibliography}
\end{document}